\documentclass[12pt]{article}
\usepackage{amssymb,amsmath,epsfig}

\begin{document}

\title{\textbf{ Models of Collapsing and Expanding Anisotropic Gravitating Source in $f(R,T)$ Theory of Gravity}}

\author{G. Abbas \thanks{abbasg91@yahoo.com} and Riaz Ahmed
\thanks{ahmedoriya@gmail.com}\\
Department of Mathematics The Islamia University\\ of Bahawalpur,
Bahawalpur, Pakistan.}
\date{}

\maketitle
\begin{abstract}
In this paper, we have formulated the exact solutions of the non-static
anisotropic gravitating source in $f(R,T)$ gravity which may lead to expansion and collapse. By assuming the no thermal conduction in gravitating source, we have determine parametric solutions in $f(R,T)$ gravity with non-static spherical geometry filled with anisotropic fluid. We have examined the range of parameter for which expansion scalar become negative and positive leading to collapse and expansion, respectively. Further, using the definition of mass function the condition for the trapped surface have been explored and it has been investigated there exists a single horizon in this case. The impact of coupling parameter $\lambda$ has been discussed in detail in both cases. For the various values of coupling parameter $\lambda$, we have plotted energy density, anisotropic pressure and anisotropic parameter in case of collapse and expansion. The physical significance of the graphs has been explained in detail.

\end{abstract}

\section{Introduction}
The gravity is the fundamental tool which is so much related to our daily life problems, it still remains the most difficult to understand and interpret interaction from all the others. The gravitational force is the most form of without any sophisticated and deep knowledge, and the first one to be tested experimentally due to the nature and apparatus needed \cite{28}. \\
Although a lot of work has been done \cite{31}-\cite{39} to explore the dynamical aspects of stars without finding the exact solutions in $f(R, T)$ gravity. The $f(R, T )$ theory of gravity has been introduced by Harko et al. \cite{2}, for the formulation of this theory they have modified the Lagrangian of general relativity as a general function of $R$ (Ricci scalar) and $T$ (trace of the stress-energy tensor). They have formulated the equations of motion by using the metric approach instead of Platini approach. It has been investigated that the importance of $T$ in the theory may be prominently observed by the exotic form of matter or phenomenological aspects of quantum gravity. The $f(R,T)$ is an explicit generalization of $f(R)$ theory, in which many cosmological and astrophysical results have discussed so far \cite{25}. But, there is a still room to study some cosmological and astronomical processes in $f(R,T)$ theory which have not yet been studied. In our present work the $f(R,T)$ model can be selected in the following form
\begin{equation}\label{m1}
f(R,T)=f_{1}(R)+f_{2}(T).
\end{equation}
Here, we take $f_{1}(R)=R$ and $f_{2}(T) =2\lambda T$
where $R$ is Ricci scalar, $\lambda$ is some positive constant and $T$ is the trace of the stress-energy tensor as already mentioned.

Recently, Zubair et al. \cite{44} analyzed the dynamical stability of cylindrically symmetric collapsing object with locally anisotropic fluid in $f(R,T)$ theory. Alves et al. \cite{23} investigated the existence of spacetime fluctuation in $f(R, T )$ and $f(R, T^{\phi})$ theories of gravity. The study of collapse and dynamics of collisionless self-gravitating systems has been carried out by the coupled collisions using the Boltzmann and Poisson equations in $f(R,T)$ gravity \cite{51,17}. Chakraborty \cite{10} have proved that unknown generalized function $f(R,T)$ can be evaluated in the closed form if this theory obeys the conservation of stress-energy tensor. Sharif and Zubair \cite{11} derived the energy conditions in $f(R,T)$ gravity which corresponds to the results of $f(R)$ gravity. Houndjo et al. \cite{3} investigated some Little Rip model in $f(R, T )$ gravity using the standard reconstruction approach. Also, they remarked that  the second law of thermodynamics remains valid for Little Rip model if the temperature inside the horizon is same that of apparent horizon.\\

Oppenheimer and Snyder \cite{8} investigated the process of collapse in 1939, they observed the contraction of inhomogeneous spherically symmetric dust cloud. This study involves the exterior and interior regions as Schwarzschild and Friedman like solutions respectively.
It has been investigated in \cite{46} that when massive stars collapse by the force of its own gravity, the final fate of such gravitational collapse is a white dwarf, neutron star, a black hole or a naked singularity. Misner and Sharp \cite{9a,9b} studied perfect fluid spherically symmetric collapse and also some authors \cite{10a}-\cite{20a} have discussed the phenomena of gravitational collapse using the dissipative and viscous fluid in general relativity. It has  been shown that one can goes beyond the general relativity, the more one has chances of admitting an uncovered singularity \cite{9}.
In fact most modified gravity theories go out of their way to avoid making any changes to gravity near massive objects. The reason for this is that there is no evidence of any odd gravitational behavior near massive objects, so all of the modified gravity theories are designed to match the standard gravity at short non-galactic distances. There are also a large number of modified gravity theories that attempt to deal with dark energy, i.e. they explain the observations that lead us to think dark energy exists as a modified gravity in a similar fashion to how modified Newtonian dynamics is supposed to do away with the need for dark matter. As far as it has proposed a modified gravity theory that consistently accounts for all observational data, might work really well for galaxies as well as for cosmology. That being said, the current dark energy and dark matter model for the Universe is not completely without its own warts so modified gravity remains a possibility, albeit less likely on current evidence than the current model.

\section{${f(R,T)}$ theory of gravity}
Harko and his collaborators \cite{2,9} proposed a generalization of the $f(R)$ theories, as $f(R, T)$ gravity.
It depends on general function of $R$ (Ricci scalar) and $T$ the trace of tensor $T_{\mu\nu}$ but in $f(R)$ theories action depends on just Ricci scalar $R$. According to the authors, the dependence of theory on $T$ (the trace of tensor $T_{\mu\nu}$) arises from quantum mechanical aspects which are usually neglected in $f(R)$ or GR theories, for instance. The full action of $f(R, T)$ gravity is in \cite{2} as  follows
\begin{eqnarray}
S=\int{d^{4}x\sqrt{-g}}(f(R,T)+L_{m}),
\end{eqnarray}
where $L_{m}$ is matter Lagrangian and $g$ is determinant of metric $g_{ab}$. Here, we choose $L_{m}=\rho,$ and above action yields
\begin{eqnarray}
G_{\mu\nu}=\frac{1}{f_{R}}\left[(f_{T}+1)T^{(m)}_{\mu\nu}-\rho g_{\mu\nu}f_{T}+\frac{f-R f_{R}}{2} g_{\mu\nu}+(\nabla_{\mu}\nabla_\nu-g_{\mu\nu}\square)f_{R}\right],
\end{eqnarray}
where $T^{(m)}_{\mu\nu}$ is matter stress-energy tensor. The spacetime in this case has the following form \\
\begin{equation}
ds^{2}=W^{2}(t,r)dt^{2}-X^{2}(t,r)dr^{2}-Y^{2}(t,r)(d\theta^{2}+sin^{2}\theta d\phi^{2}).
\end{equation}
The stress-energy tensor for anisotropic source is
\begin{equation}
{T_{uv}}^{(m)}=(\rho+P_\bot)V_{u} V_{v}-P_\bot g_{uv}+(P_r-P_\bot)X_{u} X_{v}.
\end{equation}
Here $\rho$, $V_{u}$, $X_{u}$, $P_r,~~~P_\bot$ are energy density of matter, co-moving four-velocity of the source fluid, radial four vector, radial and tangential pressures, respectively. Also, for the given line element the quantities appearing in ${T_{uv}}^{(m)}$, must satisfy \\
\begin{equation}
V^{u}=W^{-1}\delta_{0}^u, \quad V^{u}V_{u}=1,\quad  X^{u}=X^{-1}\delta_{1}^u,\quad X^{u}X_{u}=-1.
\end{equation}
The volumetric rate of expansion $\Theta$ is
\begin{equation}\label{sca}
\Theta=\frac{1}{WXY}\left({\dot{X}Y}+{2{X}\dot{Y}}\right),
\end{equation}
where $\cdot=\partial_t$ and $'=\partial_r$.
The dimensionless anisotropy is
\begin{equation}\label{a}
\Delta{a}=\frac{P_{r}-P_{\bot}}{P_{r}}.
\end{equation}
For the given line element and stress-energy tensor the set of field equation is
\begin{eqnarray}\label{4}
G_{00}&=&\frac{W^2}{f_{R}}\left[\rho+\frac{f-R f_{R}}{2}+\frac{f^{{\prime\prime}}_{R}}{X^2}-
{\dot{{f_{R}}}}\left(\frac{\dot{X}}{X}-\frac{2\dot{Y}}{Y}\right)\frac{1}{W^2}-\frac{{f_{R}^\prime}}{X^2}\left(\frac{X^\prime}{X}
-\frac{2Y^\prime}{Y}\right)\right]\\
G_{01}&=&\frac{\dot{f}_{R}^\prime}{f_{R}}\left[1-\frac{W^\prime}{W}\frac{\dot{f}_{R}}{\dot{f}_{R}^\prime}-\frac{\dot{X}}{X}\frac{f_{R}{^\prime}}{\dot{f}_{R}^\prime}\right],
\\\nonumber
G_{11}&=&\frac{X^2}{f_{R}}\left[P_r+(\rho+P_r)f_{T}-\frac{f-f_{R}}{2}
-\frac{\dot{f_{R}}}{YWX^2}\left({Y\dot{W}}-{2W\dot{Y}}\right)\right.\\
&-&\left.\frac{f_{R}^\prime}{YWX^2}\left({W^\prime}{Y}+{2Y^\prime}{W}\right)\right],
\\\nonumber
G_{22}&=&\frac{Y^2}{f_{R}}\left[p_{\bot}+(\rho+P_{\bot})f_{T}-\frac{f}{2}+\frac{R f_{R}}{2}+\frac{\ddot{f_{R}}}{W^2}-\frac{f''_{R}}{X^2}-\frac{\dot{f_{R}}}{W^2}\left(\frac{\dot{W}}{W}-\frac{\dot{X}}{X}-\frac{\dot{Y}}{Y}\right)\right.\\\label{7}
&-&\left.\frac{f_{R}^{\prime\prime}}{X^2}\left(\frac{W^\prime}{W}-\frac{X^\prime}{X}-\frac{Y^\prime}{Y}\right)
\right].
\end{eqnarray}
Misner and Sharp mass is \cite{9b}
\begin{equation}\label{8}
m(t,r)=\frac{Y}{2X^2W^2}\left(X^2{W^2}+{X^2\dot{Y^2}}-{{W^2}Y^{\prime2}}\right).
\end{equation}

By using $f(R,T)$ gravity model defined by Eq.(\ref{m1}), in Eqs.(\ref{4})-(\ref{7}), we get the system of equations in the following form
\begin{eqnarray}\nonumber
(1+\lambda)\rho-\lambda P_r-2\lambda P_{\bot}&=&\frac{1}{X^2}\left(2\frac{X^\prime}{X}\frac{Y^\prime}{Y}-\frac{Y^{\prime2}}{Y^2}-2\frac{Y^{\prime\prime}}{Y}\right)+\frac{1}{Y^2}+
\frac{1}{W^2}\left(2\frac{\dot{X}}{X}\frac{\dot{Y}}{Y}+\frac{\dot{Y^2}}{Y^2}\right),\\\\\label{f1}
G_{01}&=&0=\frac{\dot{Y^\prime}}{Y}-\frac{\dot{Y}}{Y}\frac{W^\prime}{W}-\frac{\dot{X}}{X}\frac{Y^\prime}{Y},\\\nonumber\label{f2}
\lambda \rho+(1+3\lambda)P_r+2\lambda P_{\bot}&=&\frac{1}{X^2}\left(2\frac{Y^\prime}{Y}\frac{W^\prime}{W}+\frac{Y^{\prime2}}{Y^2}\right)-\frac{1}{Y^2}
+\frac{1}{W^2}\left(2\frac{\dot{Y}}{Y}\frac{\dot{W}}{W}-\frac{\dot{Y^2}}{Y^2}-2\frac{\ddot{Y}}{Y}\right)\\\\\nonumber\label{f3}
\lambda \rho+\lambda P_r+(1+4\lambda)P_{\bot}&=&\frac{1}{X^2}\left(\frac{W^{\prime\prime}}{W}+\frac{Y^{\prime\prime}}{Y}+\frac{Y^\prime}{Y}\frac{W^\prime}{W}
-\frac{X^\prime}{X}\frac{Y^\prime}{Y}-\frac{X^\prime}{X}\frac{W^\prime}{W}\right),\\\label{f4}
&+&\frac{1}{W^2}\left(\frac{\dot{X}}{X}\frac{\dot{W}}{W}+\frac{\dot{Y}}{Y}\frac{\dot{W}}{W}-\frac{\dot{X}}{X}\frac{\dot{Y}}{Y}
-\frac{Y^{\prime\prime}}{Y}-\frac{X^{\prime\prime}}{X}\right).
\end{eqnarray}
The auxiliary solution of Eq.(\ref{f2})
\begin{equation}\label{aux}
W=\frac{\dot{Y}}{Y},\quad   X=Y^\alpha,
\end{equation}
where $\alpha$ is arbitrary constant.
Now using Eq.(\ref{sca}), we have following form of expansion scalar
\begin{eqnarray}\label{24}
\Theta=(2+\alpha)Y^{(\alpha-1)}.
\end{eqnarray}
For $\alpha>-2$ and $\alpha<-2$, we have expanding and collapsing solutions.
With the help of Eq.(\ref{aux}) mass function is given by
\begin{eqnarray}\label{25}
\frac{2m(t,r)}{Y}-1=Y^{2\alpha}-\frac{Y^{\prime2}}{Y^{2\alpha}}.
\end{eqnarray}
When $Y^\prime = Y^{2\alpha}$, there exist a trapped surface at $Y =2m$, hence $Y^\prime = Y^{2\alpha}$ is trapping condition. The trapping condition $Y^\prime = Y^{2\alpha}$ has the integral
\begin{eqnarray}\label{28}
Y^{(1-2\alpha)}_{trap}=r(1-2\alpha)+H(t),
\end{eqnarray}
where $H(t)$ appears as integration function.
Using Eqs.(\ref{aux}) and (\ref{28}), we get following explicit form of source variables
\begin{eqnarray}\nonumber
\rho&=&\frac{\left(H(t)+(r-2 \alpha  r)^{\frac{1}{1-2 \alpha }}\right)^{-2}}{(1+2\lambda)(12\lambda^2+6\lambda+1)}
+\frac{2\alpha\left(H(t)+(r-2 \alpha  r)^{\frac{1}{1-2 \alpha }}\right)^{-2(1+\alpha)}}{(r-2\alpha r)^{2}(1+2\lambda)(12\lambda^2+6\lambda+1)}\\\nonumber
&-&\frac{(r-2r\alpha)^{\frac{{4\alpha}}{1-2\alpha}}\left(H(t)+(r-2 \alpha  r)^{\frac{1}{1-2 \alpha }}\right)^{-2(1+\alpha)}}{(1+2\lambda)}\\\nonumber
&+&\frac{(r-2\alpha r)^{\frac{-1+4\alpha}{1+2\alpha}}\left(H(t)+(r-2 \alpha  r)^{\frac{1}{1-2 \alpha }}\right)^{-2(1+\alpha)}(4\alpha \lambda^2-18\lambda^2+\alpha \lambda-7\lambda-1)}{(1+2\lambda)(12\lambda^2+6\lambda+1)}\\
&+&\frac{(1+2\alpha)\left(H(t)+(r-2 \alpha  r)^{\frac{1}{1-2 \alpha }}\right)^{4\alpha}(8\alpha \lambda^2+12\lambda^2+2\alpha \lambda+6\lambda+1)}{(1+2\lambda)(12\lambda^2+6\lambda+1)},\\\nonumber
\end{eqnarray}
\begin{eqnarray}\nonumber
P_r&=&\frac{\left(H(t)+(r-2 \alpha  r)^{\frac{1}{1-2 \alpha }}\right)^{2(-1+\alpha)}(1+2\alpha)(12\lambda^2-2\alpha \lambda+6\lambda+1)}{(1+2\lambda)(12\lambda^2+6\lambda+1)}\\\nonumber
&+&\frac{\left(H(t)+(r-2 \alpha  r)^{\frac{1}{1-2 \alpha }}\right)^{-2(1+\alpha)}(r-2 \alpha  r)^{\frac{2}{1-2 \alpha }}(6\lambda^2+5\lambda+1)}{(1+2\lambda)(12\lambda^2+6\lambda+1)r^2(-1+2\alpha)}\\\nonumber
&+&\frac{(r-2\alpha r)^{\frac{-1+4\alpha}{1-2\alpha}}\left(H(t)+(r-2 \alpha  r)^{\frac{1}{1-2 \alpha }}\right)^{-2(1+\alpha)}(r-2\alpha r)^{\frac{-1+4\alpha}{1-2\alpha}}}{(1+2\lambda)(12\lambda^2+6\lambda+1)}\\\nonumber
&\times& \left[-4(\alpha^2\lambda+6\alpha\lambda^2)-(r-2 \alpha  r)^{\frac{1}{1-2 \alpha }}(12\alpha\lambda+6\lambda+1)\right]
\\&+&\frac{\left(H(t)+(r-2 \alpha  r)^{\frac{1}{1-2 \alpha }}\right)^{-2}}{(1+2\lambda)},\\\nonumber
P_{\bot}&=&\frac{\alpha  (1-2 \lambda ) (r-2 \alpha  r)^{\frac{4 \alpha }{1-2 \alpha }} \left(H(t)+(r-2 \alpha  r)^{\frac{1}{1-2 \alpha }}\right)^{-2 (\alpha +1)}}{12 \lambda ^2+6 \lambda +1}\\
&-&\frac{(1-2 \alpha )^2 (2 \alpha +1) (2 \lambda +1) r^2 \left(H(t)+(r-2 \alpha  r)^{\frac{1}{1-2 \alpha }}\right)^{2 (\alpha -1)}}{\left(12 \lambda ^2+6 \lambda +1\right) (r-2 \alpha  r)^2}.
\end{eqnarray}
\section{Generating Solutions}
For various possible values of $\alpha$, we discuss the nature of solutions:
\subsection{Collapse solution $\alpha=-\frac{5}{2}$}
When the value of expansion scalar is negative, we have gravitational collapse, so Eq.(\ref{24}), implies that $\Theta<0$, if $\alpha<-2$, for the convenience we take $\alpha=-\frac{5}{2}$ and the condition $Y^\prime=Y^{2\alpha}$, provides $Y^\prime=Y^{-5}$, which further integrated to
\begin{eqnarray}\label{col1}
Y_{trap}=\left(6r+h_{1}(t)\right)^\frac{1}{6}.
\end{eqnarray}
Here $h_1(t)$ is integration function.
Using Eqs.(\ref{aux}), (\ref{col1}) in Eqs.(\ref{f1})-(\ref{f4}), with some tedious algebra, we obtain the following explicit form of matter variables
\begin{eqnarray}
\mathfrak{}\rho&=&-\frac{5\lambda(1+4\lambda)}{\left(6r+h_{1}(t)\right)^\frac{7}{6}(1+2\lambda)(12\lambda^2+6\lambda+1)}\\\nonumber
&+&\frac{1}{\left(6r+h_{1}(t)\right)^\frac{1}{3}(1+2\lambda)},\\
P_r&=&-\frac{5 \lambda }{\left(24 \lambda ^3+24 \lambda ^2+8 \lambda +1\right) (6r+h_{1}(t))^{7/6}}\\\nonumber
&-&\frac{1}{(2 \lambda +1) \sqrt[3]{6r+h_{1}(t)}},\\
P_{\bot}&=&-\frac{5(1+2\lambda)}{2\left(6r+h_{1}(t)\right)^\frac{7}{6}(12\lambda^2+6\lambda+1)}.\\\nonumber
\end{eqnarray}
The mass function given in Eq.(\ref{24}), becomes
\begin{eqnarray}
m_{1}(r,t)=\frac{1}{2} \sqrt[6]{6r+h_{1}(t)}.
\end{eqnarray}
The dimensionless parameter $\Delta{a}$ from Eq.(\ref{a}) takes the form
\begin{eqnarray}
\Delta{a}&=&\frac{-5-10\lambda(1+2\lambda)+\left(6r+h_{1}(t)\right)^\frac{5}{6}(2+6\lambda+12\lambda^2)}
{10\lambda+2\left(6r+h_{1}(t)\right)^\frac{5}{6}(1+6\lambda+12\lambda^2)}.
\end{eqnarray}

\begin{figure}
\begin{center}
\includegraphics[width=120mm]{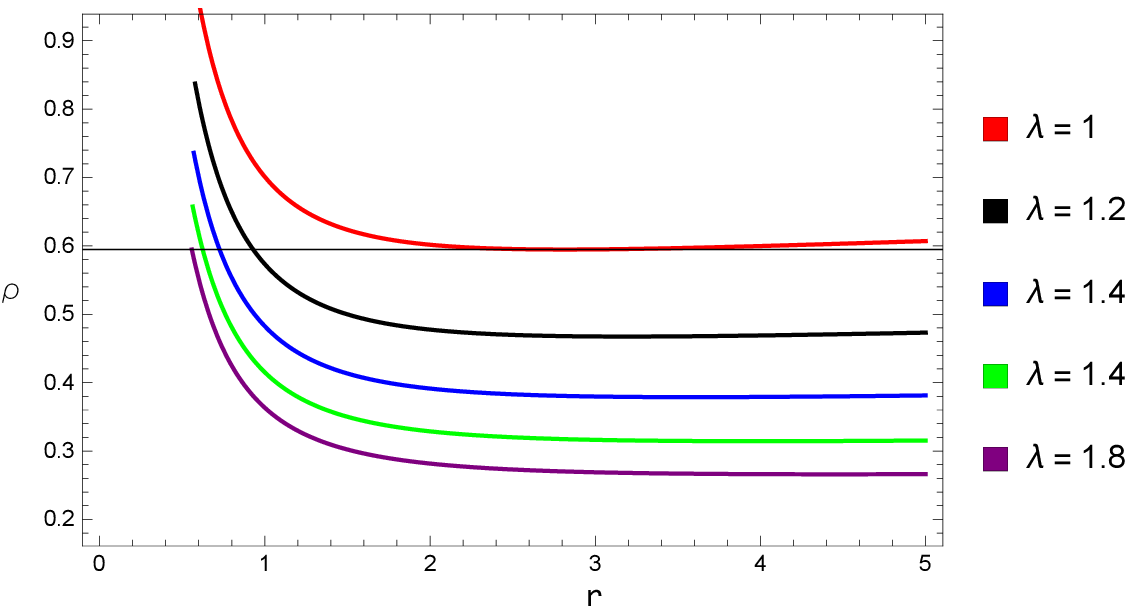}
\caption{This graph shows the variation of $\rho$ with respect to $r$ for the various values of $\lambda$ and $h_{1}(t)$ = 1}
\end{center}
\end{figure}

\begin{figure}
\begin{center}
\includegraphics[width=120mm]{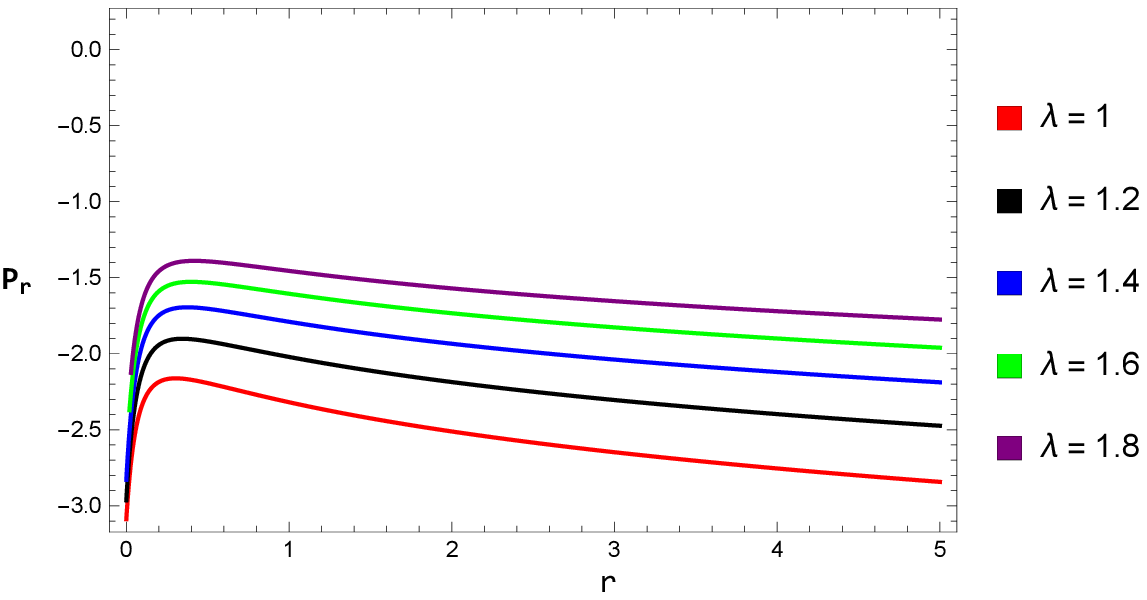}
\caption{This graph shows the variation of $P_r$ with respect to $r$ for the various values of $\lambda$ and $h_{1}(t)$ = 1}
\end{center}
\end{figure}
\begin{figure}
\begin{center}
\includegraphics[width=120mm]{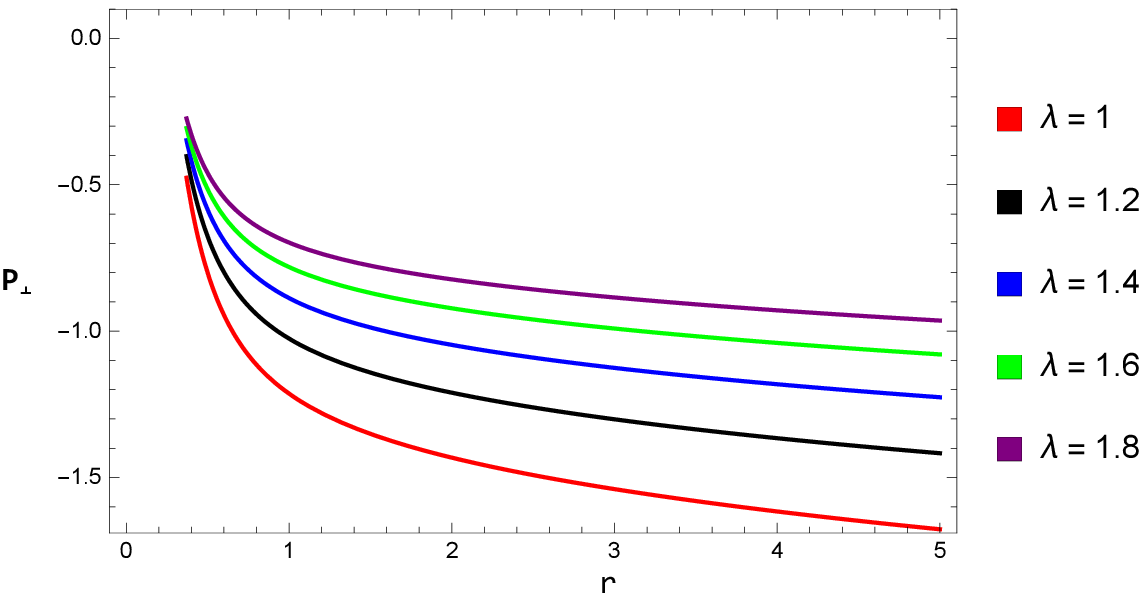}
\caption{This graph shows the variation of $P_{\bot}$ with respect to $r$ for the various values of $\lambda$ and $h_{1}(t)$ = 1}
\end{center}
\end{figure}
\begin{figure}
\begin{center}
\includegraphics[width=120mm]{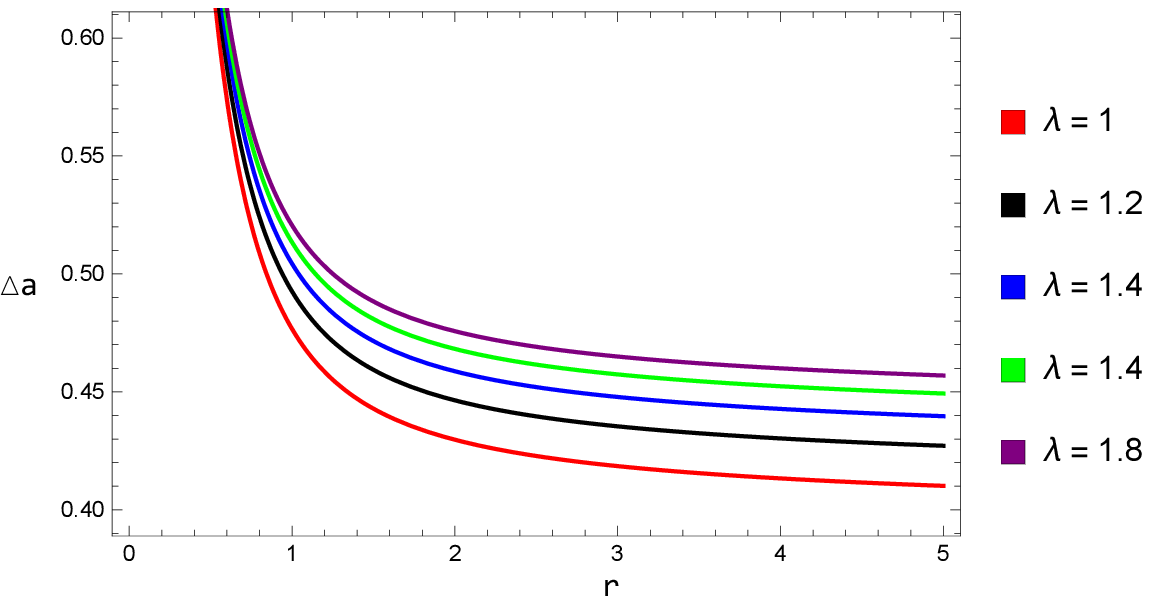}
\caption{This graph shows the variation of $\triangle a$ with respect to $r$ for the various values of $\lambda$ and $h_{1}(t)$ = 1}
\end{center}
\end{figure}

In the above expressions $h_{1}(t)$ is an arbitrary function of time profile $t$, by taking $h_{1}(t)=1$, we have analyzed the results. By choosing $\alpha = -\frac{5}{2}$, we get $\Theta<0$ and energy density remains positive and decreasing function of $r$ the graphical behavior of $\rho$ with various values of $\lambda$ is shown in \textbf{Fig.1}. The radial pressure increases firstly and then decreases continuously with respect to radius at different values of  $\lambda$ as shown in \textbf{Fig.2}, but transverse pressure is decreasing with respect to radius as shown in \textbf{Fig.3}. In account of this we can say pressure is minimum when the value of $\lambda$ is minimum as shown in \textbf{2} and \textbf{3}. The maximum value of anisotropy occurs near the center of the sphere, so the anisotropic parameter $\Delta a$ attains maximum value near the center and its values decreases when $r$ increases with various values of $\lambda$ it is given in \textbf{Fig.4}.
\subsection{Expansion with $\alpha=\frac{3}{2}$}
When expansion scalar attains positive values, we have expanding solution, so Eq.(\ref{24}), implies that $\Theta>0$, if $\alpha>-2$, for the convenience we take $\alpha=\frac{3}{2}$ and assume that
\begin{eqnarray}\label{exp1}
Y=(r^2+r^{2}_{0})^{-1}+h_{2}(t),
\end{eqnarray}
where $h_{2}(t)$ is integration function and $r_{0}>0$. For simplicity, we take $F(t,r)=1+h_{2}(t)(r^2+r^{2}_{0})$ and $Y=\frac{F}{(r^2+r^{2}_{0})}$, then Eqs.(\ref{aux}), (\ref{exp1}) with Eqs.(\ref{f1})-(\ref{f4}), with simplification, we obtain the following explicit form of matter variables
\begin{eqnarray}\nonumber
\rho&=&\frac{14\lambda^2+7\lambda+1}{(1+2\lambda)(12\lambda^2+6\lambda+1)}\left[-\frac{8r^2(r^2+r^{2}_{0})}{F^5}+\frac{4(3r^2-r^{2}_{0})}{F^4}-\frac{4F}{(r^2+r^{2}_{0})}-\frac{(r^2+r^{2}_{0})^2}{F^2}\right]\\\nonumber
&+&\frac{8\lambda^2+2\lambda}{(1+2\lambda)(12\lambda^2+6\lambda+1)}\left[\frac{6}{(r^2+r^{2}_{0})F}-\frac{12r^2(r^2+r^{2}_{0})}{F^5}+\frac{(3r^2-r^{2}_{0})(r^2-r^{2}_{0})}{F^4}\right]\\
&+&\frac{\lambda}{(12\lambda^2+6\lambda+1)}\left[\frac{8r^2(r^2+r^{2}_{0})}{F^5}+\frac{(r^2+r^{2}_{0})^2}{F^2}+\frac{4F}{(r^2+r^{2}_{0})}\right],
\end{eqnarray}
\begin{eqnarray}\nonumber
P_r&=&\frac{\lambda(1+6\lambda)}{(1+2\lambda)(12\lambda^2+6\lambda+1)}\left[\frac{8r^2(r^2+r^{2}_{0})}{F^5}+\frac{4(3r^2-r^{2}_{0})(r^2+r^{2}_{0})}{F^4}+\frac{(r^2+r^{2}_{0})^2}{F^2}+\frac{4F}{r^2+r^{2}_{0}}\right]\\\nonumber
&+&\frac{6\lambda^2+5\lambda+1}{(1+2\lambda)(12\lambda^2+6\lambda+1)}\left[\frac{8r^2(r^2+r^{2}_{0})}{F^5}++\frac{(r^2+r^{2}_{0})^2}{F^2}+\frac{4F}{r^2+r^{2}_{0}}\right]\\
&-&\frac{2\lambda}{(1+2\lambda)(12\lambda^2+6\lambda+1)}\left[\frac{(r^2+r^{2}_{0})^2\left(3r^4-r^{2}_{0}(1+r^{2}_{0})+r^2(2r^{2}_{0}-9)\right)}{F^5}+\frac{6F}{(r^2+r^{2}_{0})}\right],
\end{eqnarray}
\begin{eqnarray}\nonumber
P_{\bot}&=&\frac{\lambda}{(12\lambda^2+6\lambda+1)}\left[\frac{8r^2(r^2+r^{2}_{0})}{F^5}-\frac{4(3r^2-r^{2}_{0})(r^2+r^{2}_{0})}{F^4}+\frac{4F}{(r^2+r^{2}_{0})}+\frac{(r^2+r^{2}_{0})^2}{F^2}\right]\\\nonumber
&-&\frac{1}{(12\lambda^2+6\lambda+1)}\left[\frac{6}{(r^2+r^{2}_{0})F}-\frac{12r^(-4+F)(r^2+r^{2}_{0})}{F^5}-\frac{r^{2}_{0}(r^2+r^{2}_{0})}{F^4}\right]\\
&-&\frac{1+2\lambda}{(12\lambda^2+6\lambda+1)}\left[\frac{8r^2(r^2+r^{2}_{0})}{F^5}+\frac{(r^2+r^{2}_{0})^2}{F^2}+\frac{4F}{(r^2+r^{2}_{0})}\right].
\end{eqnarray}
The anisotropic parameter and mass function are given by
\begin{eqnarray}\nonumber
\triangle a&=&\frac{4\lambda^2\left(\frac{8r^2(r^2+r^{2}_{0})}{F}-4(3r^2-r^{2}_{0})(r^2+r^{2}_{0})+(r^2+r^{2}_{0})F^2+4F^5\right)}{(1+2\lambda)\left(4(3r^2-r^{2}_{0})(r^2+r^{2}_{0})\lambda+\frac{6(1+2\lambda)F^5}{(r^2+r^{2}_{0})}+\frac{r^2(r^2+r^{2}_{0})(2r^{2}_{0}-9)(1+2\lambda)}{F}\right)}\\\nonumber
&+&\frac{(r^2+r^{2}_{0})(4\lambda^2+2\lambda+1)\left(\frac{3r^4-r^{2}_{0}(1+r^{2}_{0})}{F}+\frac{6F^5}{(r^2+r^{2}_{0})^2}+\frac{r^2(2r^{2}_{0})-9}{F}\right)}{(1+2\lambda)\left(4(3r^2-r^{2}_{0})(r^2+r^{2}_{0})\lambda+\frac{6(1+2\lambda)F^5}{(r^2+r^{2}_{0})}+\frac{r^2(r^2+r^{2}_{0})(2r^{2}_{0}-9)(1+2\lambda)}{F}\right)}\\\nonumber
&+&\frac{(8\lambda^2+6\lambda+1)\left(\frac{8r^2(r^2+r^{2}_{0})}{F}+(r^2+r^{2}_{0})F^2+4F^5\right)}{(1+2\lambda)\left(4(3r^2-r^{2}_{0})(r^2+r^{2}_{0})\lambda+\frac{6(1+2\lambda)F^5}{(r^2+r^{2}_{0})}+\frac{r^2(r^2+r^{2}_{0})(2r^{2}_{0}-9)(1+2\lambda)}{F}\right)},\\
\end{eqnarray}
\begin{eqnarray}
m_{2}(t,r)&=&\frac{1}{2}\left(\frac{F}{(r^2+r^{2}_{0})}-\frac{4r^2}{(r^2+r^{2}_{0})^4F^4}+\frac{F^4}{(r^2+r^{2}_{0})^4}\right).
\end{eqnarray}
\begin{figure}
\centering
\includegraphics[width=120mm]{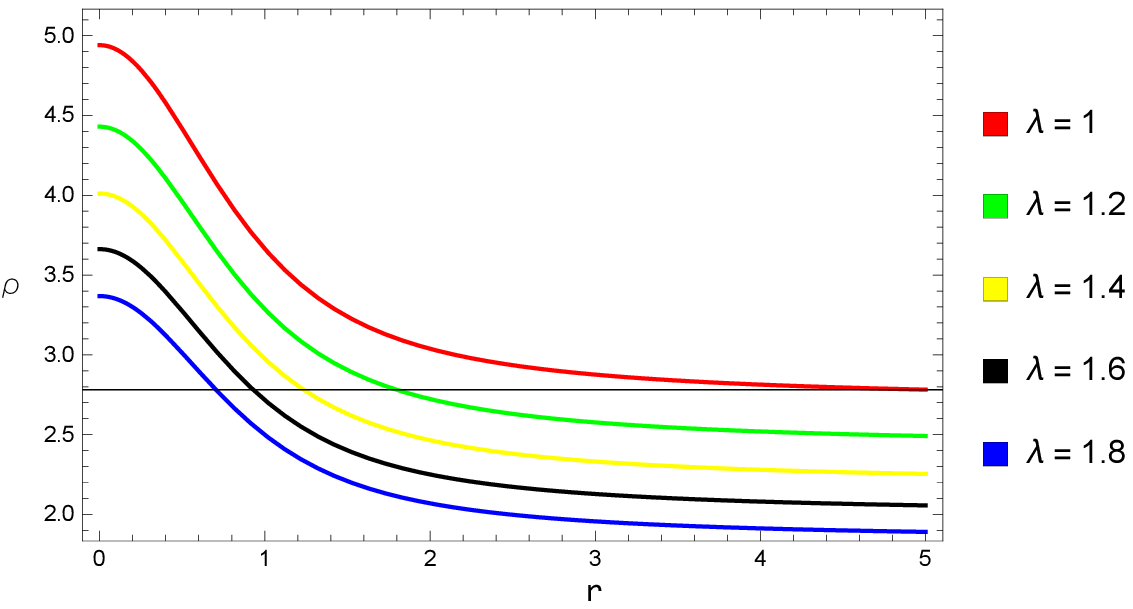}
\caption{This graph shows the variation of $\rho$ with respect to $r$ for the various values of $\lambda$ and $h_{2}(t)$ = 1}
\end{figure}
\begin{figure}
\begin{center}
\includegraphics[width=120mm]{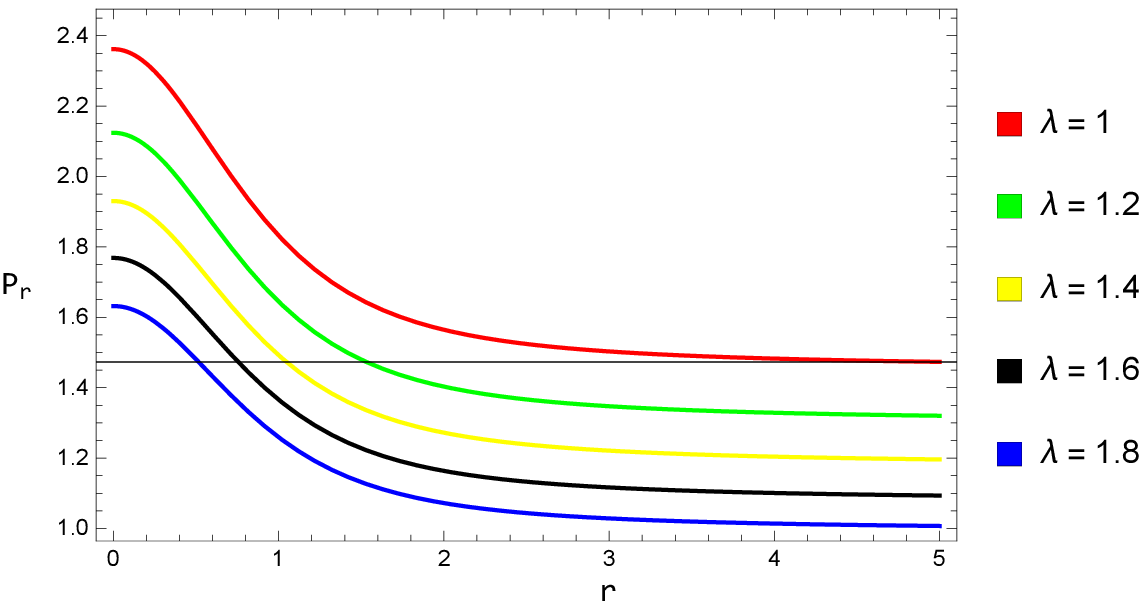}
\caption{This graph shows the variation of $P_r$ with respect to $r$ for the various values of $\lambda$ and $h_{2}(t)$ = 1}
\end{center}
\end{figure}
\begin{figure}
\begin{center}
\includegraphics[width=120mm]{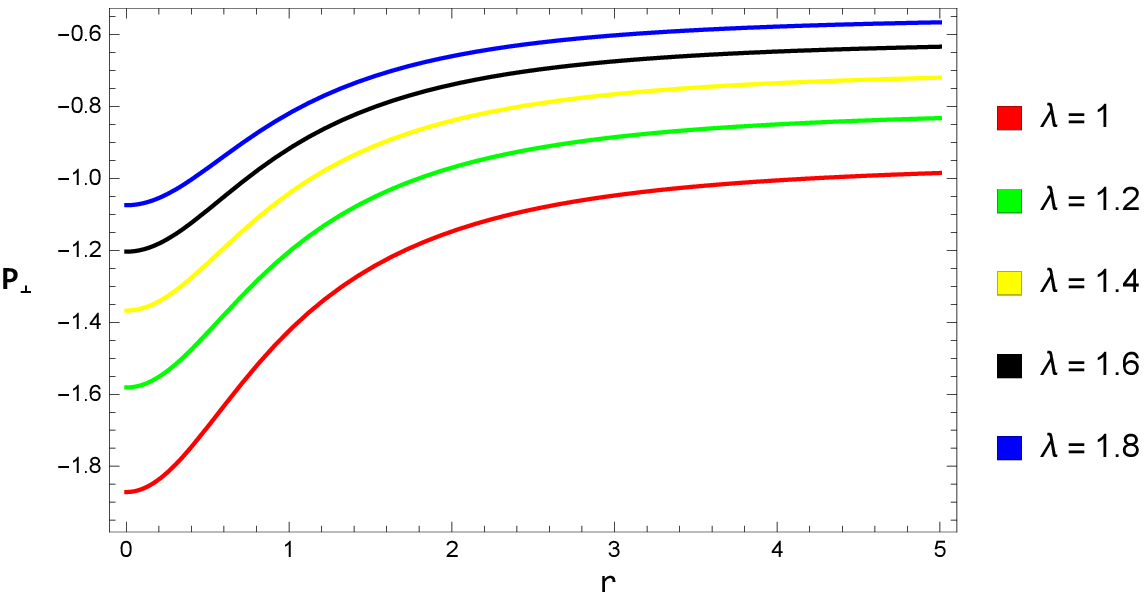}
\caption{This graph shows the variation of $P_{\bot}$ with respect to $r$ for the various values of $\lambda$ and $h_{2}(t)$ = 1}
\end{center}
\end{figure}
\begin{figure}
\begin{center}
\includegraphics[width=120mm]{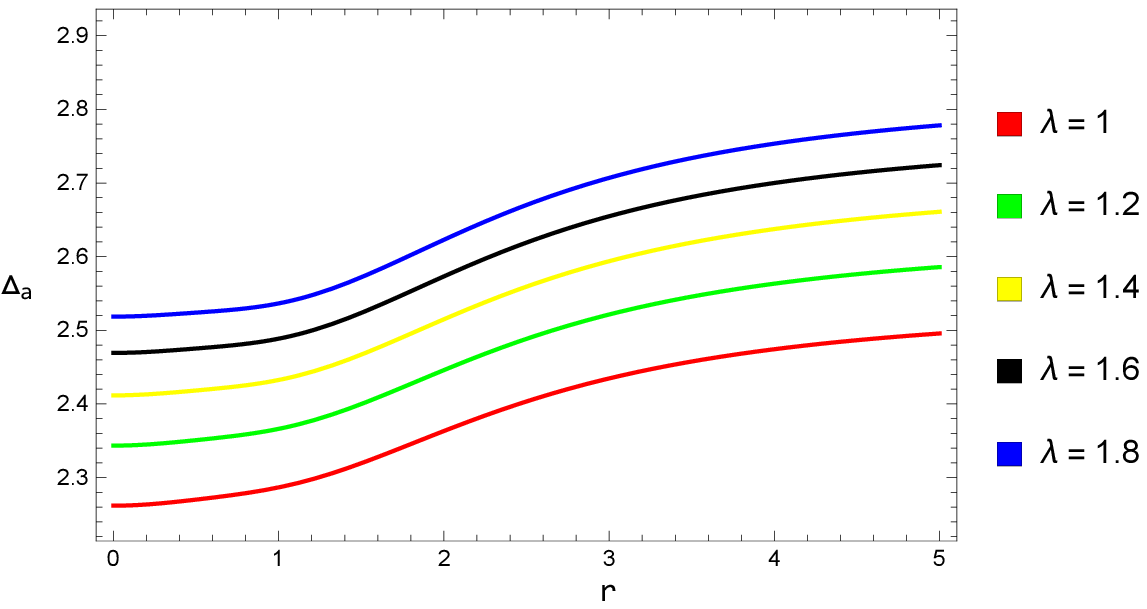}
\caption{This graph shows the variation of $\triangle a$ with respect to $r$ for the various values of $\lambda$ and $h_{2}(t)$ = 1}
\end{center}
\end{figure}.\\
 In this case, when $\alpha = \frac{3}{2}$ we get $\Theta>0$ and energy density remains finitely positive for selected arbitrary time profile $t$, by fixing the values of $\alpha$, the variation is given to $\lambda$ under such restrictions the density $\rho$  can be observed decreasing function shown in \textbf{Fig.$5$}. The radial pressure decreases continuously with respect to radius at different values of $\lambda$, but transverse pressure is increasing with respect to radius as shown in \textbf{Figs.} \textbf{6} and \textbf{7} respectively. In account of this we can say pressure has reverse effects as observed in previous case. It is observed that the anisotropy is increasing function of $r$ but with respect to the value of $\lambda$ anisotropy is going to be maximum as the value of $\lambda$ is being increased as shown in \textbf{Fig.8}.
\section{Conclusion}
Motivated by $f(R,T)$ theory of gravity formulated by Harko et al. \cite{2}, a lot of work related to cosmology and stability of dynamics of collapsing stellar system has been done in recent years \cite{12}-\cite{c32}. This theory has a vide range of cosmological and astrophysical applications in the era of modern physics. According to the available observation data, our universe is in the phase of accelerating expansion, to explain the physical significance of this phenomena a number of modifications to GR have been proposed. The $f(R,T)$ theory of gravity is one has been the center of attention of researchers in current era this type of theories seems to provide a capacity of working successfully for dark matter. The conformal relation of $f(R,T)$ to GR with a self-interacting scalar field has been discussed by Zubair et. al.\cite{22}.\\
Here, we have developed the generating solutions to collapse and expansion of fluid sphere in f(R,T) theory of gravity. For the interior matter distribution, the collapse solution yields a unique trapped surface. It has been investigated that during gravitational contraction, the phase transition would occur for the massive stellar system, for instance most of condensed matter configuration transit to a $\pi$-meson condensed state. The gravitational collapse is a highly dissipative process and a lot amount of heat energy is released during gravitational collapse Herrera at. el. \cite{54}. In order to model the inhomogeneous cosmological solutions Collins \cite{53} have explored the non-static expanding solutions. Also, the inclusion of anisotropic stress in the fluid source are much important, the influence of non-zero anisotropic parameter $\delta a$ on the late late time evaluation of universe with non-homogeneous background has been explored by Barrow et. al. \cite{55}. Due to the valid selection of  $R(t,r)$ and $\alpha$, one can obtain anisotropy interconnection which collapses or expands by Glass \cite{56}. In this paper, we extend the work of Glass \cite{56} to the $f(R,T)$ theory of gravity.\\

In this paper interior solution for anisotropic fluids have been discussed in detail, which are being used in modeling of anisotropic stars in the context of modified theory of gravity $f(R,T)$. Using the auxiliary form of of the metric functions, we have determined the trapping conditions for fluid sphere in $f(R,T)$ gravity. The resulting solutions have been classified as collapsing and expanding depending on the the nature of scalar expansion. The matter density, radial and transverse pressures, anisotropic parameter and mass function have been calculated in the context of $f(R,T)$ theory of gravity. For the collapse solution when $\alpha = -\frac{5}{2}$) the density decreases as the value of $\lambda$ increase as shown in  \textbf{Fig. \textbf{1}}. The radial pressure $P_r$ and matter density $\rho$ have maximum values at center and it decreases from center to the surface of star. It has been observed that the anisotropy will be directed outward when $P_r>P_{\bot}$ this gives that $\triangle a >0$ as observed graphically in \textbf{Figs}.\textbf{4} and \textbf{8}. It is found in fig \textbf{4} that anisotropy decreases with the increase in radius.  \\
Further, the expansion of gravitating source would occur when $\alpha = \frac{3}{2}$, and $\Theta$ the expansion scalar is positive. In this case matter density decreases as shown in \textbf{Fig.} \textbf{5}. The radial/transverse pressures and anisotropic parameter with various values of $\lambda$ have reverse behavior as compared to the case of gravitational collapse.

\end{document}